\documentclass[twocolumn,showpacs,prl]{revtex4}
\usepackage{amssymb}
\usepackage{amsmath}
\usepackage{graphicx}
\usepackage{dcolumn}
\usepackage{bm}

\begin{document}

\title{State-independent experimental test of quantum contextuality in an
indivisible system}
\author{C. Zu$^{1}$, Y.-X. Wang$^{1}$, D.-L. Deng$^{1,2}$, X.-Y. Chang$^{1}$%
, K. Liu$^{1}$, P.-Y. Hou$^{1}$, H.-X. Yang$^{1}$, L.-M. Duan$^{1,2}$}
\affiliation{$^{1}$Center for Quantum Information, IIIS, Tsinghua University, Beijing,
China}
\affiliation{$^{2}$Department of Physics, University of Michigan, Ann Arbor, Michigan
48109, USA}

\begin{abstract}
We report the first state-independent experimental test of quantum
contextuality on a single photonic qutrit (three-dimensional system), based on a recent theoretical
proposal [Yu and Oh, Phys. Rev. Lett. 108, 030402 (2012)]. Our experiment spotlights quantum contextuality in its
most basic form, in a way that is independent of either the state or the
tensor product structure of the system.
\end{abstract}

\maketitle

Contextuality represents a major deviation of quantum theory from
classical physics \textbf{\cite{1,2}}. Non-contextual realism is a pillar of the familiar worldview of classical
physics. In a non-contextual world, observables have pre-defined values,
which are independent of our choices of measurements. Non-contextuality
plays a role also in the derivation of Bell's inequalities, as the property
of local realism therein can be seen as a special form of non-contextuality, 
where the independence of the measurement context is enforced by the
no-signalling principle \textbf{\cite{2a,2b,8,9}}. In an attempt to save the
non-contextuality of the classical worldview, non-contextual hidden variable
theories have been proposed as an alternative to quantum mechanics. In these
theories, the outcomes of measurements are associated to hidden variables,
which are distributed according to a joint probability distribution.
However, the celebrated Kochen-Specker theorem \textbf{\cite{1,2,2a,2b}}
showed that non-contextual hidden variable theories are incompatible with
the predictions of quantum theory. The original Kochen-Specker theorem is
presented in the form of a logical contradiction, which is conceptually
striking, but experimentally unfriendly: the presence of unavoidable
experimental imperfections motivated a debate on whether or not the
non-contextual features highlighted by Kochen-Specker theorem can be
actually tested in experiments \textbf{\cite{10,11}}. As a result of the
debate, new Bell-type inequalities have been proposed in the recent years,
with the purpose of pinpointing the contextuality of quantum mechanics in an
experimentally testable way. These inequalities are generally referred to as
the \emph{KS\ inequalities} \textbf{\cite{8}}. Violation of the KS\
inequalities confirms quantum contextuality and rules out the non-contextual
hidden variable theory. Different from the Bell inequality tests, violation
of the KS\ inequality can be achieved independently of the state of quantum
systems \textbf{\cite{1,2,8,9}}, showing that the conflict between quantum
theory and non-contextual realism resides in the structure of quantum
mechanics instead of particular quantum states. The KS inequalities have
been tested in experiments for two qubits, using ions \textbf{\cite{3}},
photons \textbf{\cite{4,4a}}, neutrons \textbf{\cite{5}}, or an ensemble
nuclear magnetic resonance system \textbf{\cite{6}}. A single qutrit
represents the simplest system where it is possible to observe conflict
between quantum theory and non-contextual realistic models \textbf{\cite%
{7,9,9a,9b}}. A recent experiment has demonstrated quantum contextuality for
photonic qutrits in a particular quantum state \textbf{\cite{7}}, based on a
version of the KS inequality proposed by Klyachko, Can, Binicioglu, and
Shumovsky \textbf{\cite{9a}}.

A state-independent test of quantum contextuality for a single qutrit, in
the spirit of the original KS\ theorem, is possible but complicated as one
needs to measure many experimental configurations \textbf{\cite{2b,9,9b}}. A
recent theoretical work by Yu and Oh proposes another version of the KS\
inequality, which requires to measure $13$ variables and $24$ of their pair
correlations \textbf{\cite{9}}. This is a significant simplification
compared with the previous KS inequalities for single qutrits, and the
number of variables cannot be further reduced as proven recently by Cabello
\textbf{\cite{12}}. Our experiment confirms quantum contextuality in a
state-independent fashion using the Yu-Oh version of the KS inequality for
qutrits represented by three distinctive paths of single photons. The
maximum violation of this inequality by quantum mechanics is only $4\%$
beyond the bound set by the non-contextual hidden variable theory, so we
need to accurately control the paths of single photons in experiments to
measure the $13$ variables and their correlations for different types of
input states. We have achieved a violation of the KS\ inequality by more
than five standard deviations for all the nine different states that we
tested.

For a single qutrit with basis vectors $\left\{ \left\vert 0\right\rangle
,\left\vert 1\right\rangle ,\left\vert 2\right\rangle \right\} $, we detect
projection operators to the states $i\left\vert 0\right\rangle +j\left\vert
1\right\rangle +k\left\vert 2\right\rangle $ specified by the $13$ unit
vectors $\left( i,j,k\right) $ in Fig. 1. The $13$ projectors have
eigenvalues either $0$ or $1$. In the hidden variable theory, the
corresponding observables are assigned randomly with values $0$ or $1$
according to a (generally unknown) joint probability distribution. When two
states are orthogonal, the projectors onto them commute, and the
corresponding observables are called compatible, which means that they can
be measured simultaneously. Non-contextuality means that the assignment of
values to an observable should be independent of the choice of compatible
observables that are measured jointly with it. For instance, $z_{1}$ in Fig.
1 should be assigned the same value in the correlators $z_{1}z_{2}$ and $%
z_{1}y_{1}^{\pm }$ for each trial of measurement. For each observable $%
b_{i}\in \left\{ z_{\mu },y_{\mu }^{\pm },h_{\alpha },\mu =1,2,3;\alpha
=0,1,2,3\right\} $ defined in Fig. 1, we introduce a new variable $%
a_{i}\equiv 1-2b_{i}$, which takes values of $\pm 1$. For the $13$
observables $a_{i}$ with two outcomes $\pm 1$, it is shown in Ref. \textbf{%
\cite{9}} that they satisfy the inequality%
\begin{equation}
\mathop{\displaystyle \sum }\limits_{i}a_{i}-\frac{1}{4}\mathop{%
\displaystyle \sum }\limits_{\left\langle i,j\right\rangle }a_{i}a_{j}\leq 8,
\end{equation}%
where $\left\langle i,j\right\rangle $ denotes all pairs of observables that
are compatible with each other. There are $24$ compatible pairs among all
the $13\times 13$ combinations, and a complete list of them is given in
Table 1 for the corresponding operator correlations. The inequality (1)\ can
be proven either through an exhaustive check of all the possible $2^{13}$
value assignments of $a_{i}$ $\left( i=1,2,\cdots ,13\right) $ or by a more
elegant analytic argument as shown in Ref. \textbf{\cite{9}}. In quantum
theory, each $a_{i}$ corresponds to an operator $A_{i}$ with eigenvalues $%
\pm 1$ in quantum mechanics. In the hidden variable theory, the value $a_{i}$
corresponds to a random variable $A_{i}$, and the different values are
distributed according to a (possibly correlated) joint probability
distribution. Hence, for the hidden variable theory the expectation values
of $A_{i}$ must satisfy the inequality
\begin{equation}
\mathop{\displaystyle \sum }\limits_{i}\left\langle A_{i}\right\rangle -%
\frac{1}{4}\mathop{\displaystyle \sum }\limits_{\left\langle
i,j\right\rangle }\left\langle A_{i}A_{j}\right\rangle \leq 8.
\end{equation}%
which follows by taking the average of (1) over the joint probability
distribution of the values $a_{i}$. On the other hand, quantum theory gives
a different prediction: From the definition $A_{i}\equiv I-2B_{i}$, where $%
B_{i}$ is the projection operator to the $13$ states in Fig. 1, we find that
$S=\mathop{\displaystyle \sum }\limits_{i}A_{i}-\frac{1}{4}%
\mathop{\displaystyle \sum }\limits_{\left\langle i,j\right\rangle
}A_{i}A_{j}\equiv \frac{25}{3}I$, where $I$ is the unity operator. Hence,
for any state of the system, quantum theory predicts the inequality $%
\left\langle S\right\rangle =25/3\nleq 8$, which violates the inequality (2)
imposed by the non-contextual hidden variable theory and rules out any
non-contextual realistic model.

Since the quantum mechanical prediction $\langle S\rangle =25/3$ is close to
the upper bound $\langle S\rangle \leq 8$ set by the non-contextual realism,
we need to achieve accurate control in experiments to violate the inequality
(2). Yu and Oh also derived another simpler inequality in Ref. \textbf{\cite%
{9}} by introducing an additional assumption (as proposed in the original
KS\ proof \textbf{\cite{1,2b}}) that the algebraic structure of compatible
observables is preserved at the hidden variable level, that is, that the
value assigned to the product (or sum) of two compatible observables is
equal to the product (or sum) of the values assigned to these observables.
Under this assumption, it is shown in \textbf{\cite{9}} that
\begin{equation}
\mathop{\displaystyle \sum }\limits_{\alpha =0,1,2,3}\left\langle
B_{h_{\alpha }}\right\rangle \leq 1
\end{equation}%
for non-contextual hidden variable theory, while quantum mechanically $%
\mathop{\displaystyle \sum }\limits_{\alpha =0,1,2,3}B_{h_{\alpha }}\equiv
\frac{4}{3}I$, and thus $\mathop{\displaystyle \sum }\limits_{\alpha
=0,1,2,3}\left\langle B_{h_{\alpha }}\right\rangle =4/3>1$. The inequality
(3) is more amenable to experimental tests than Eq. (2), as it requires only
four measurement settings. However, conceptually it is weaker than Eq. (2)
due to the additional assumption required for its proof. Our experiment
achieves significant violation of both the inequalities (2) and (3).

To experimentally test the inequalities (2) and (3), first we prepare a
single photonic qutrit through the spontaneous parametric down conversion
(SPDC) setup shown in Fig. 2. The SPDC process generates correlated
(entangled) photon pairs, and through detection of one of the photons by a
detector D0, we get a heralded single-photon source on the other output
mode. This photon is then split by two polarizing beam splitters (PBS) into
three spatial modes that represent a single photonic qutrit. Through control
of the wave plates before the PBS\ and for the pump light, we can prepare
any state for this photonic qutrit.

The state of the qutrit is then detected by three single-photon detectors
D1-D3. To measure the observables $A_{i}$ and their correlations, we use the
setup shown in Fig. 1 based on cascaded Mach-Zehnder interferometers. The
wave plates HWP3 and HWP4 in the interferometers can be tilted to fine tune
the phase difference between the two arms. To stabilize the relative phase,
the whole interferometer setup is enclosed in a black box. The detectors
D1-D3 measure projections to three orthogonal states in the qutrit space,
which always correspond to mutually compatible observables. By tuning the
half wave plates HWP5 and HWP6 in Fig. 1, we can choose these projections so
that they give a subset of the $13$ projection operators $B_{i}$. A detector
click (non-click) then means assignment of value $1$ ($0$) to the
corresponding observable $B_{i}$ (or equivalently, assignment of $-1$ $%
\left( +1\right) $ to the observable $A_{i}$). The coincidence between the
detectors measures the correlation. The detailed configurations of the wave
plates to measure different correlations are summarized in section 1 of the
supplementary information. Due to the photon loss, sometimes our photonic
qutrit does not yield a click in the detectors D1-D3, even though we
registered a heralding photon at the detector D0. To take this into account,
we discard the events when none of the detectors D1-D3 fires, in the same
way as it was done in Ref. \textbf{\cite{7}}. The use of this post-selection
technique opens up a detection efficiency loophole, and we need to assume
that the events selected out by the photonic coincidence is an unbiased
representation of the whole sample (\emph{fair-sampling assumption}).

We have measured all the expectation values in the inequality (2) and (3)
for different input states. Table 1 summarizes the measurement results for a
particular input state $\left\vert s\right\rangle =\left( \left\vert
0\right\rangle +\left\vert 1\right\rangle +\left\vert 2\right\rangle \right)
/\sqrt{3}$ in equal superposition of the three basis-vectors. The
theoretical values in the quantum mechanical case are calculated using the
Born rule with the ideal state $\left\vert s\right\rangle $. Each of the
experimental correlations is constructed from the joint probabilities $%
P\left( A_{i}=\pm 1;A_{j}=\pm 1\right) $ registered by the detectors. As an
example to show the measurement method, in section 2 of the supplementary
information, we give detailed data for the registered joint probabilities
under different measurement configurations, which together fix all the
correlations in Table 1. The expectation value $\left\langle
B_{i}\right\rangle $ (or $\left\langle A_{i}\right\rangle \equiv
1-2\left\langle B_{i}\right\rangle $) is directly determined by the relative
probability of the photon firing in the corresponding detector. From the
data summarized in Table 1, we find both of the inequalities (2) and (3) are
significantly violated in experiments, in agreement with the quantum
mechanics prediction and in contradiction with the non-contextual realistic
models. Even the tough inequality (2)\ is violated by more than five times
the error bar (standard deviation).

To verify that the inequalities (2) and (3) are experimentally violated
independently of the state of the system, we have tested them for different
kinds of input states. The set of states tested include the three
basis-vectors $\left\{ \left\vert 0\right\rangle ,\left\vert 1\right\rangle
,\left\vert 2\right\rangle \right\} $, the two-component superposition
states $\left\{ \left( \left\vert 0\right\rangle +\left\vert 1\right\rangle
\right) /\sqrt{2},\left( \left\vert 0\right\rangle +\left\vert
2\right\rangle \right) /\sqrt{2},\left( \left\vert 1\right\rangle
+\left\vert 2\right\rangle \right) /\sqrt{2}\right\} $, the three-component
superposition state $\left\vert s\right\rangle $, and two mixed states $\rho
_{8}=\left( \left\vert 0\right\rangle \left\langle 0\right\vert +\left\vert
2\right\rangle \left\langle 2\right\vert \right) /2$ and $\rho _{9}=\left(
\left\vert 0\right\rangle \left\langle 0\right\vert +\left\vert
1\right\rangle \left\langle 1\right\vert +\left\vert 2\right\rangle
\left\langle 2\right\vert \right) /3\equiv I/3$. The detailed configurations
of the wave plates to prepare these different input states are summarized in
section 1 of the supplementary information. To generate the mixed states, we
first produce photon pairs entangled in polarization using the type-I phase
matching in the BBO\ crystal \textbf{\cite{14}}. After tracing out the idler
photon by the detection at D0, we get a mixed state in polarization for the
signal photon, which is then transferred to a mixed qutrit state represented
by the optical paths through the PBS. For various input states, we measure
correlations of all the observables in the inequality (2) and the detailed
results are presented in section 3 of the supplementary information.
Although the expectation values $\left\langle A_{i}\right\rangle $ and the
correlations $\left\langle A_{i}A_{j}\right\rangle $ strongly depend on the
input states, the inequalities (2) and (3) are state-independent and
significantly violated for all the cases tested in experiments. In Fig. 3,
we present the measurement outcomes of these two inequalities for nine
different input states. The results violate the boundary set by the
non-contextual hidden variable theory and are in excellent agreement with
quantum mechanics predictions.

In this work, we have observed violation of the KS inequalities (2) and (3)
for a single photonic qutrit, which represents the first state-independent
experimental test of quantum contextuality in an indivisible quantum system.
The experiment confirmation of quantum contextuality in its most basic form,
in a way that is independent of either the state or the tensor product
structure of the system, sheds new light on the contradiction between
quantum mechanics and non-contextual realistic models.

\textbf{Acknowledgement} This work was supported by the National Basic
Research Program of China (973 Program) 2011CBA00300 (2011CBA00302) and the
NSFC Grant 61033001. DLD and LMD acknowledge in addition support from the
IARPA MUSIQC program, the ARO and the AFOSR MURI program.

\begin{figure}[tbp]
\includegraphics[width=8cm,height=4cm]{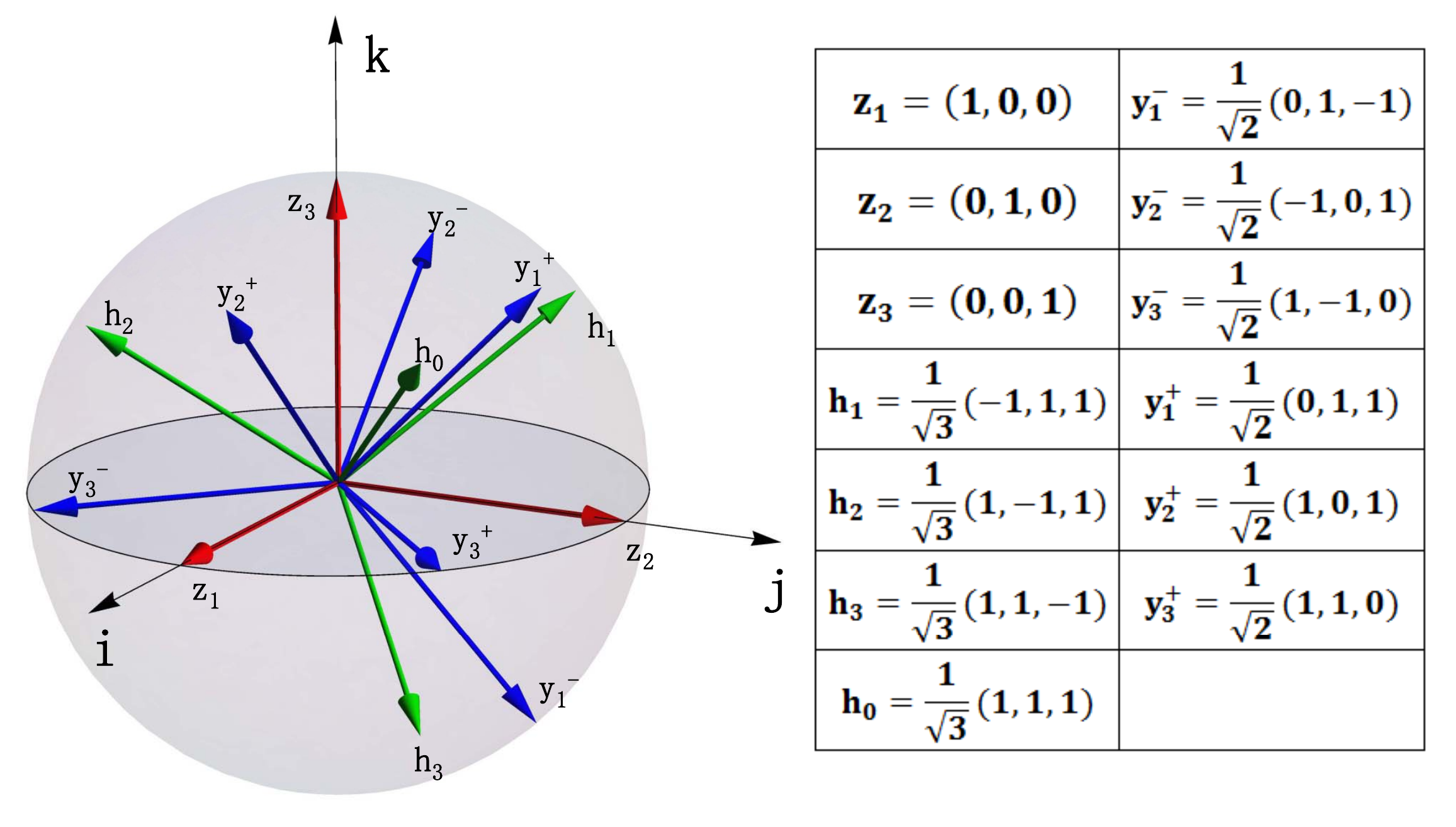}
\caption[Fig. 1 ]{ Illustration of the $13$ unit-vectors that describe the
superposition coefficients of $13$ corresponding qutrit states. The KS
inequality proposed in Ref. \protect\cite{9} requires to detect projection
operators onto these $13$ states and their correlations. }
\end{figure}

\begin{widetext}

\begin{figure}[tbp]
\includegraphics[width=18cm,height=6cm]{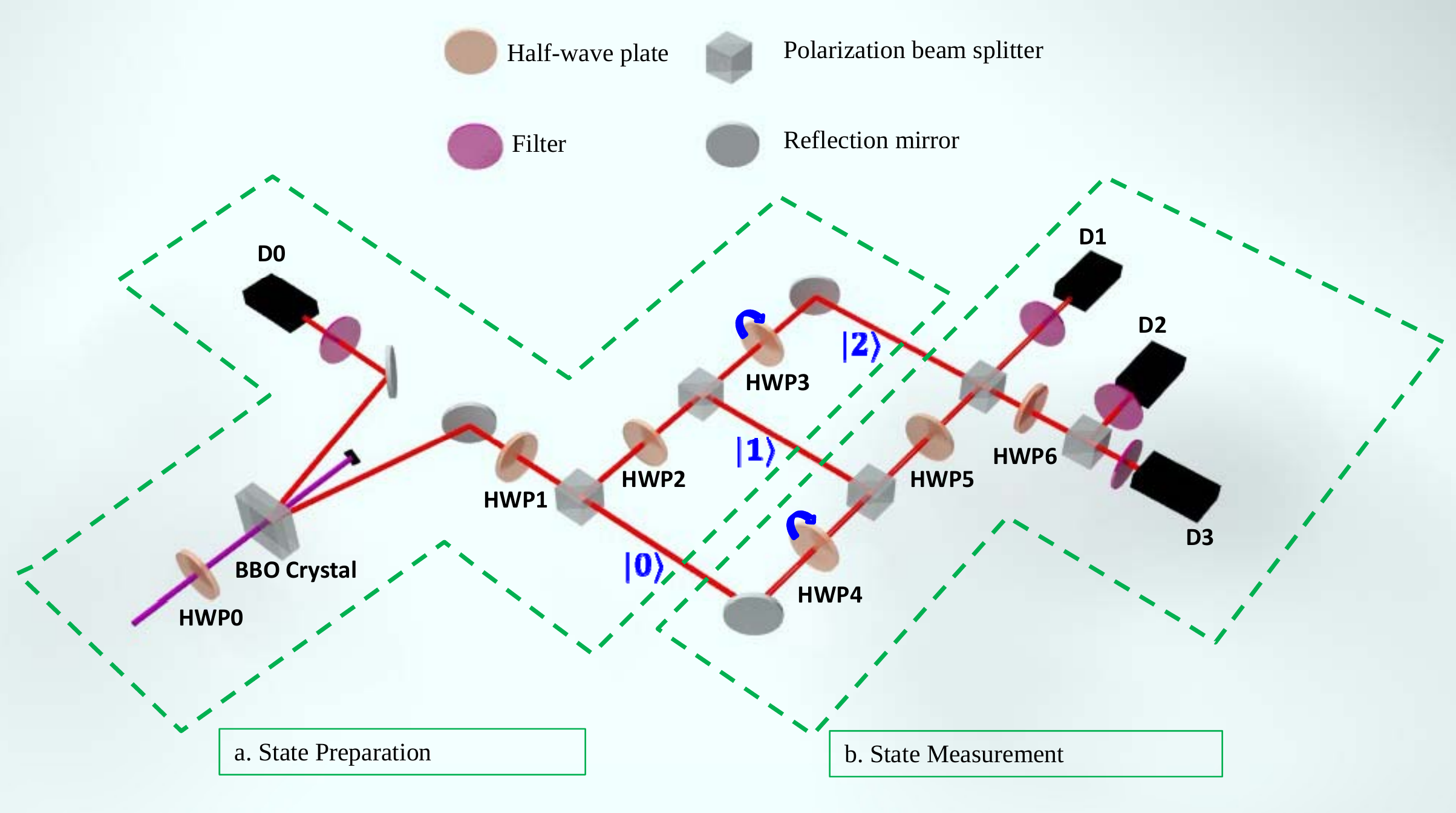}
\caption[Fig. 3 ]{ Illustration of the experimental setup to detect the KS inequalities. The setup in the box (a) is for
state preparation of a single photonic qutrit. Ultrafast laser pulses (with a
repetition rate of $76$ MHz) at the wavelength of $400$ nm from a frequency
doubled Ti:sapphire laser pump two joint beta-barium-borate (BBO)\ crystals,
each of $0.6$ mm depth with perpendicular optical axis, to generate
correlated (entangled) photon pairs at the wavelength of $800$ nm. With registration of
a photon-count at the detector D0, we get a heralded single photon source in the other output mode.
This photon is split by two polarizing beam splitters (PBS) into three optical paths,
representing a single photonic qutrit. By adjusting the angle of the half wave plates (HWP1 and HWP2),
we can control the superposition coefficients of this qutrit state. The setup in box (b) is for measurement
of the qutrit state along compatible projections to three orthogonal states. By tuning the wave plates
(HWP5 and HWP6), we choose these projection operators to be along the directions specified by the $A_i$ operators to
measure the correlations of the compatible $A_i$. The wave plates HWP3 and HWP4 are used to balance the Mach-Zender
interferometers and can be tilted for fine tuning of the relative phase. }
\end{figure}

\begin{figure}[tbp]
\includegraphics[width=18cm,height=7cm]{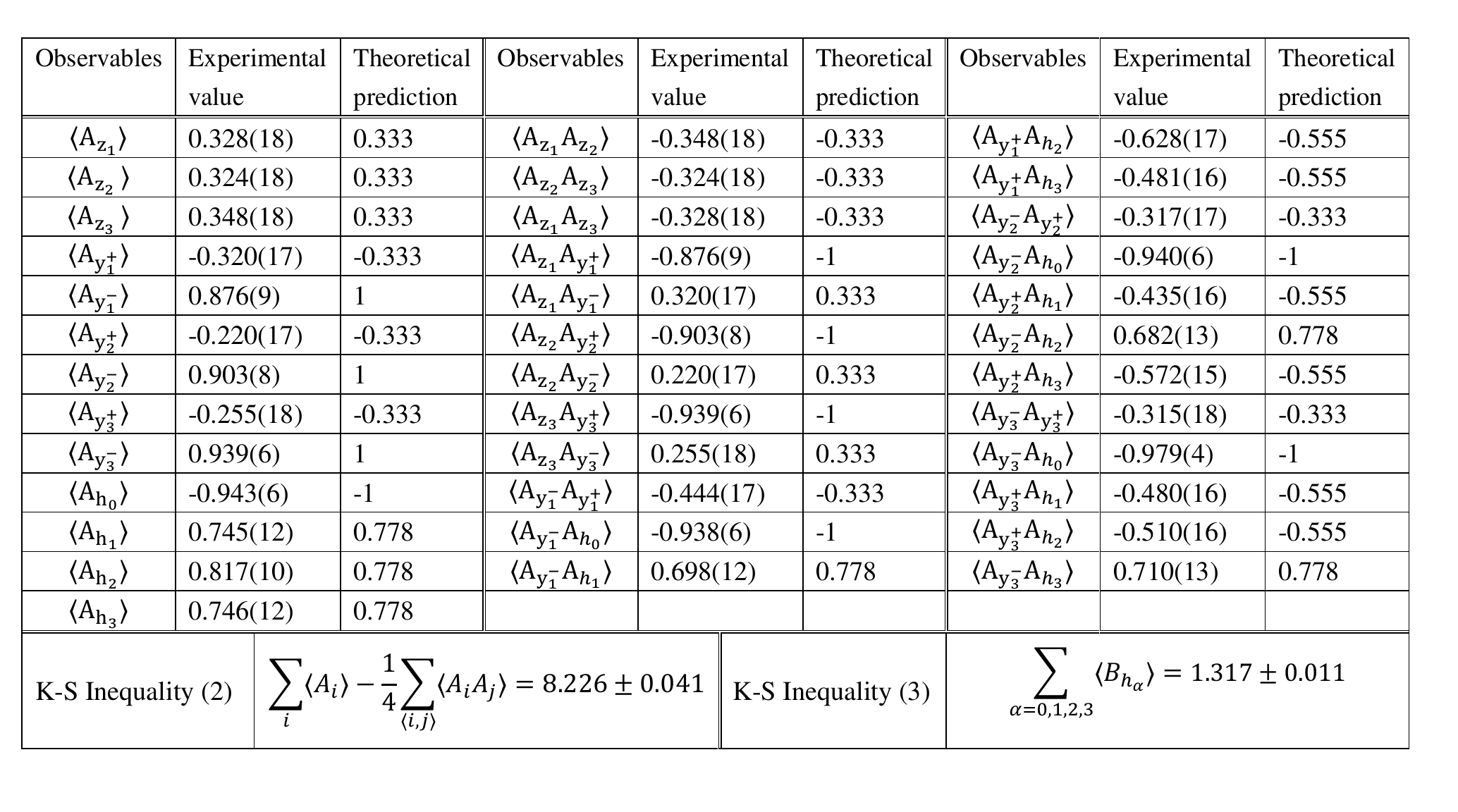}
\caption[Table. 1 ]{ Table  1: (A) The measured expectation values $\left\langle A_{i}\right\rangle $ and
the correlations $\left\langle A_{i}A_{j}\right\rangle $ for all the compatible pairs under
a particular input state $\left\vert s\right\rangle =\left( \left\vert
0\right\rangle +\left\vert 1\right\rangle +\left\vert 2\right\rangle \right)
/\sqrt{3}$. For the experimental values, the numbers in the bracket represent the statistical error associated
with the photon detection under the assumption of a Poissonian distribution
for the photon counts, for instance, $\left\langle A_{z_1}\right\rangle=0.328(18)\equiv0.328\pm0.018$. Both of the inequalities (2) and (3) are significantly violated by
the experimental data.}
\end{figure}

\end{widetext}

\begin{figure}[tbp]
\includegraphics[width=8cm,height=5cm]{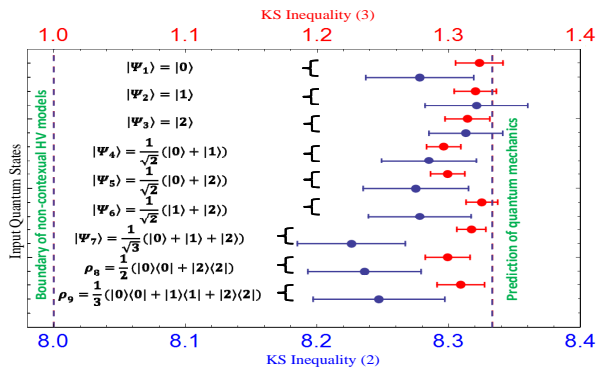}
\caption[Fig. 4 ]{ (A) The measurement results for the two inequalities (2)
(shown by the blue label) and (3) (shown by the red label) under different
types of input states. The left side dash line specifies the upper bounds
imposed by any non-contextual hidden variable models while the right side
dash line corresponds to the quantum mechanical prediction under the ideal
input states. The error bars account for the statistical error associated
with the photon detection.}
\end{figure}

\newpage

\section{Supplementary information: State-independent experimental test of
quantum contextuality in an indivisible system}

This supplementary information gives the detailed configurations and data
for the experimental test of state-independent quantum contextuality on a
single photonic qutrit. In Sec. I, we first give the configurations of the
wave plates to prepare different input states for a single photonic qutrit,
and then summarize the configurations of the experiment to measure all the
observables and their correlations in the KS\ inequalities. In Sec. II, we
show the method to calculate the correlations form the measured joint
probabilities, using a particular input state as an example. In Sec. III, we
give the detailed data of the correlation measurements for the other eight
input states that are not present in the main manuscript.


\section{Configurations of the wave plates for state preparation and
measurement}

\begin{table}[tbp]
\includegraphics[width=8.5cm,height=6cm]{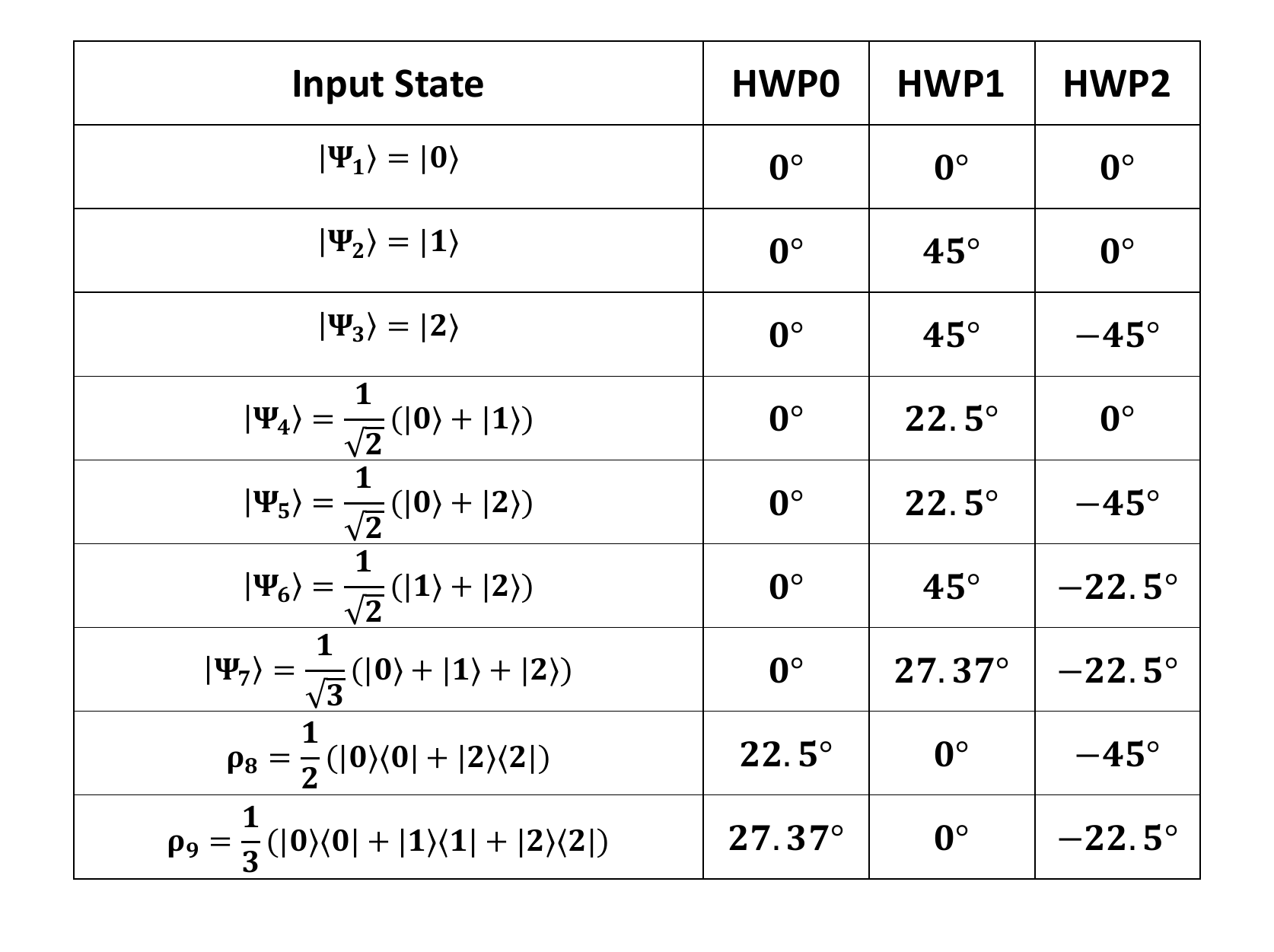}
\caption[Supplementary Table. 1 ]{Angles of half-wave plates (HWP) to
prepare different states for a single photonic qutrit.}
\end{table}

To demonstrate that the violation of the KS\ inequalities is independent of
the state of the system, we need to prepare different types of input states
for the single photonic qutrit. This is achieved by adjusting the angles of
three half wave plates HWP0, HWP1, and HWP2 in the experimental setup shown
in Fig. 2 of the manuscript. To prepare pure input state, the polarization
of the pumping laser is set to $|V\rangle $ (vertically polarized) by the
HWP0. With the type-I phase matching in the BBO\ nonlinear crystal, the
generated signal and idler photons are both in the polarization state $%
|H\rangle $. After the heralding measurement of the idler photon, the
polarization of the signal photon is rotated by the HWP1 and HWP2. The
polarization beam splitter (PBS) transmits the photon when it is in $%
|H\rangle $ polarization and reflects it when it is in $|V\rangle $
polarization. The half-wave plate with angle $\theta $ transfers the
polarization basis-states $\left\vert H\right\rangle $ and $\left\vert
V\right\rangle $ by the formula $\left\vert H\right\rangle \rightarrow \cos
\left( 2\theta \right) \left\vert H\right\rangle +\sin \left( 2\theta
\right) \left\vert V\right\rangle $ and $\left\vert V\right\rangle
\rightarrow \cos \left( 2\theta \right) \left\vert V\right\rangle -\sin
\left( 2\theta \right) \left\vert H\right\rangle $. To prepare an arbitrary
input state $c_{0}|0\rangle +c_{1}|1\rangle +c_{2}|2\rangle $, the angle of
the HWP1 sets the branching ratio $c_{0}/\sqrt{c_{1}^{2}+c_{2}^{2}}$, and
the angle of the HWP2 then determines $c_{1}/c_{2}$. For the seven pure
input states in the experiment, the corresponding angles of the HWP1 and
HWP2 are listed in Table. I.

It is more striking to see that the KS\ inequalities are violated even for
completely mixed states. To prepare a mixed state for the signal photon, we
rotate the polarization of the pumping laser to $\left( |H\rangle +|V\rangle
\right) /\sqrt{2}$ by setting the angle of HWP0 at $22.5^{o}$. The output
state for the signal and the ideal photon after the BBO\ crystal is a
maximally entangled one with the form $|\Psi \rangle _{si}=\left( |HH\rangle
+e^{i\varphi }|VV\rangle \right) /\sqrt{2}$, where $\varphi $ is a relative
phase of the two polarization components. After the heralding measurement of
the idler photon, the state of the signal photon is described by the reduced
density matrix $(|H\rangle \left\langle H\right\vert +|V\rangle \left\langle
V\right\vert )/2$. If we set the HWP1 and HWP2 respectively at the angle of $%
0^{o}$ and $45^{o}$, this polarization mixed state is transferred to the
qutrit mixed state $\rho _{8}=(|0\rangle \left\langle 0\right\vert
+|2\rangle \left\langle 2\right\vert )/2$ as shown in Table 1. To prepare
the mixed state $\rho _{9}$ (the most noisy qutrit state), we set the HWP0
at the angle $27.37^{o}$ and the density operator for the signal photon
right after the BBO\ crystal becomes $(|H\rangle \left\langle H\right\vert
+2|V\rangle \left\langle V\right\vert )/3$. When we set the HWP1 and HWP2
respectively at $0^{o}$ and $22.5^{o}$, the photonic qutrit is described by
the state $\left[ |0\rangle \left\langle 0\right\vert +|1\rangle
\left\langle 1\right\vert +|2\rangle \left\langle 2\right\vert +\left(
e^{i\phi }|1\rangle \left\langle 2\right\vert +H.c.\right) \right] /3$,
where the relative phase $\phi =0$ in the ideal case. However, we randomly
tilt the HWP3 in this experiment which sets the phase $\phi $ to a random
value. After average over many experimental runs to measure the correlation,
the effective state for the qutrit is described by $\rho _{9}=(|0\rangle
\left\langle 0\right\vert +|1\rangle \left\langle 1\right\vert +|2\rangle
\left\langle 2\right\vert )/3=I/3$, the completely mixed state.

\begin{table}[tbp]
\includegraphics[width=8.5cm,height=6cm]{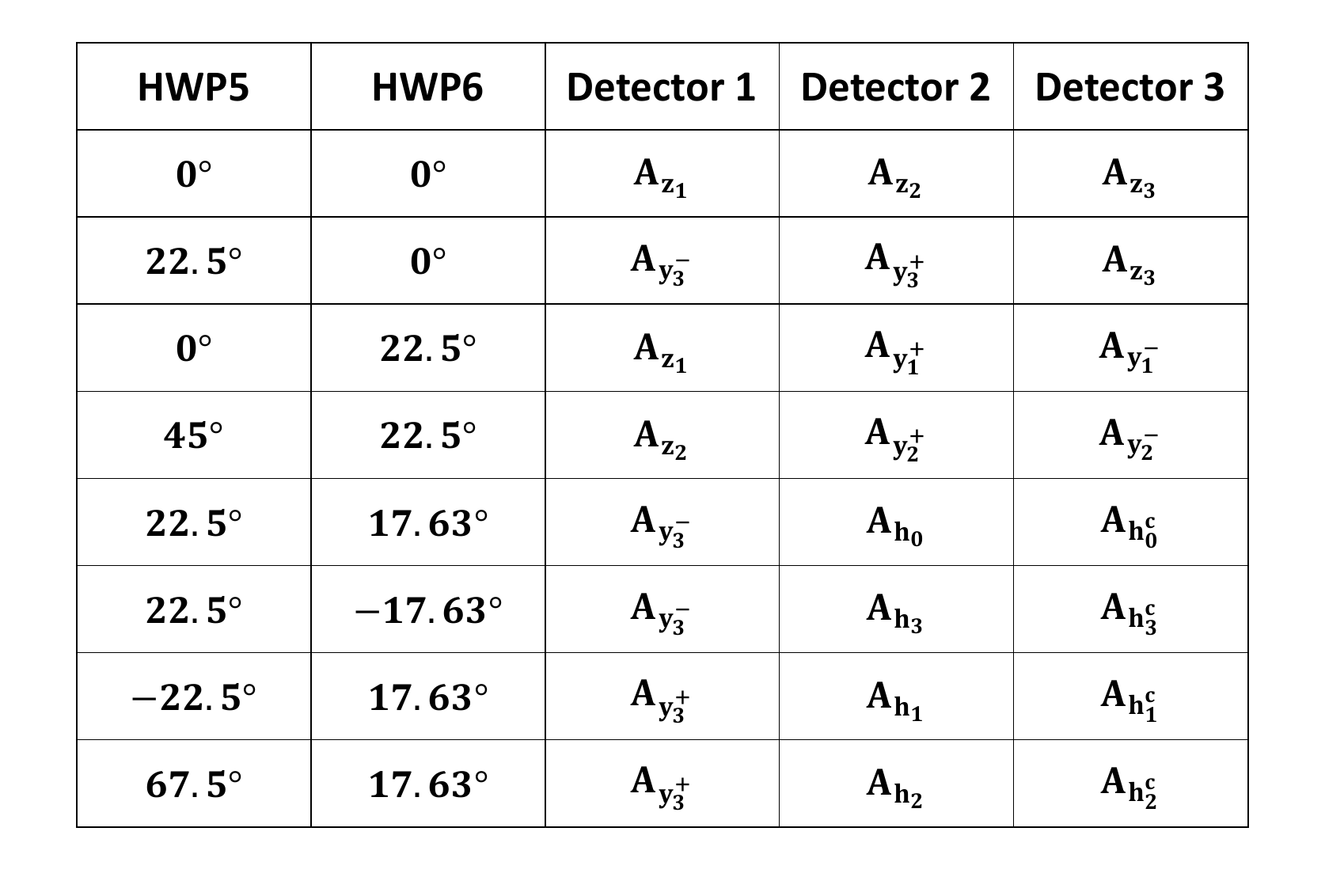}
\caption[Supplementary Table. 2 ]{Angles of half-wave plates (HWP5 and HWP6
in Fig. 1 of the main manuscript) to measure correlations of different
compatible observables, where $A_{h_{0}^{c}}$, $A_{h_{1}^{c}}$, $%
A_{h_{2}^{c}}$, and $A_{h_{3}^{c}}$\ correspond, respectively, to the
projection operators onto the states $\left\vert h_{0}^{c}\right\rangle
=(|0\rangle +|1\rangle -2|2\rangle )/\protect\sqrt{6}$, $\left\vert
h_{1}^{c}\right\rangle =(-|0\rangle +|1\rangle -2|2\rangle )/\protect\sqrt{6}
$, $\left\vert h_{2}^{c}\right\rangle =(|0\rangle -|1\rangle -2|2\rangle )/%
\protect\sqrt{6}$, and $\left\vert h_{3}^{c}\right\rangle =(|0\rangle
+|1\rangle +2|2\rangle )/\protect\sqrt{6}$. }
\end{table}

To detect the KS inequalities, we need to measure the $13$ observables $%
\left\langle A_{i}\right\rangle $ and $24$ compatible combinations of their
pair-wise correlations $\left\langle A_{i}A_{j}\right\rangle $. By rotating
the angles of the HWP5 and HWP6, We choose the measurement bases so that the
photon count at the detectors D1, D2, or D3 correspond to a measurement of
the compatible combinations of the projection operators $A_{i}$. In Table
II, we list the angles of the HWP5 and HWP6 and the corresponding operators
detected by the single-photon detectors D1, D2, and D3. With these
configurations of the wave plates, we read out the $13$ expectation values $%
\left\langle A_{i}\right\rangle $ and $16$ of their compatible pair-wise
correlations. The other $8$ compatible correlations $\left\langle A_{y_{\mu
}^{\pm }}A_{h_{\alpha }}\right\rangle $ ($\mu =1,2;\alpha =0,1,2,3$) are
obtained from the correlations $\left\langle A_{y_{3}^{\pm }}A_{h_{\alpha
}}\right\rangle $ with an exchange of the basis-vectors $|2\rangle
\leftrightarrow |0\rangle $ or $|2\rangle \leftrightarrow |1\rangle $. So,
to measure $\left\langle A_{y_{\mu }^{\pm }}A_{h_{\alpha }}\right\rangle $,
we use the same configurations as specified by the last four rows of Table II
and exchange the basis-vectors $|2\rangle \leftrightarrow |0\rangle $ (or $%
|1\rangle $) in the input state through an appropriate rotation of the HWP1
and HWP2.

\section{Calculation of correlations from the measured joint probabilities}

The outcomes for observables $A_{i}$ are either $+1$ or $-1$, depending on
whether there is a photon click (or no click) in the corresponding photon
detector. The correlation $\left\langle A_{i}A_{j}\right\rangle $ of the
compatible observables $A_{i}$ and $A_{j}$ is constructed from the four
measured joint probabilities $P(A_{i}=\pm 1,A_{j}=\pm 1)$, whereas the
latter is read out from coincidence of the single-photon detectors through
the relation
\begin{eqnarray}
\langle A_{i}A_{j}\rangle  &=&P(A_{i}=1,A_{j}=1)+P(A_{i}=-1,A_{j}=-1)  \notag
\\
-P(A_{i} &=&1,A_{j}=-1)-P(A_{i}=-1,A_{j}=1).
\end{eqnarray}%
All the events need to be heralded by the single-photon detector D0. So the
joint probability $P(A_{i}=-1,A_{j}=+1)$ corresponds to the coincidence rate
$\left\langle D0,Di\right\rangle $ of the detectors D0 and Di, normalized by
the total coincidence $\left\langle D0,D1\right\rangle +\left\langle
D0,D2\right\rangle +\left\langle D0,D3\right\rangle $. Similar expressions
hold for $P(A_{i}=-1,A_{j}=1)$\ and $P(A_{i}=+1,A_{j}=+1)$. The joint
probability $P(A_{i}=-1,A_{j}=-1)$ is proportional to the three-photon
coincidence rate $\left\langle D0,Di,Dj\right\rangle $, which is smaller
than the two-photon coincidence rate by about four orders of magnitude in
our experiment. So the probability $P(A_{i}=-1,A_{j}=-1)$ is significantly
less than the error bars for the other joint probabilities and negligible in
calculation of the correlation $\langle A_{i}A_{j}\rangle $ by Eq. (1).

In table III, we list the measured joint probabilities for all the compatible
pairs $A_{i}$ and $A_{j}$, taking the superposition state $|\Psi _{7}\rangle
=|s\rangle =(|0\rangle +|1\rangle +|2\rangle )/\sqrt{3}$ as an example for
the input. The corresponding correlations $\left\langle
A_{i}A_{j}\right\rangle $ are calculated using Eq. (1). The number in the
bracket represent the error bar on the last (or last two) digits associated
with the statistical error in the photon counts.

\section{Results of correlation measurements and KS\ inequalities for other
input states}

In the main manuscript, we have given the measured expectation values $%
\left\langle A_{i}\right\rangle $ and the correlations $\left\langle
A_{i}A_{j}\right\rangle $ for a particular input state $|\Psi _{7}\rangle
=|s\rangle =(|0\rangle +|1\rangle +|2\rangle )/\sqrt{3}$. We have tested the
KS inequalities for $9$ different input states, ranging from the simple
basis vectors, to the superposition states, and to the most noisy mixed
states. For all the input states, we have observed significant violation of
the KS inequalities for single photonic qutrits. In this section, we list
the measured expectation values $\left\langle A_{i}\right\rangle $ and the
correlations $\left\langle A_{i}A_{j}\right\rangle $ for the other $8$ input
states (shown in Table IV, V, VI, and VII).

\begin{widetext}

\begin{table}[tbp]
\includegraphics[width=18cm,height=15cm]{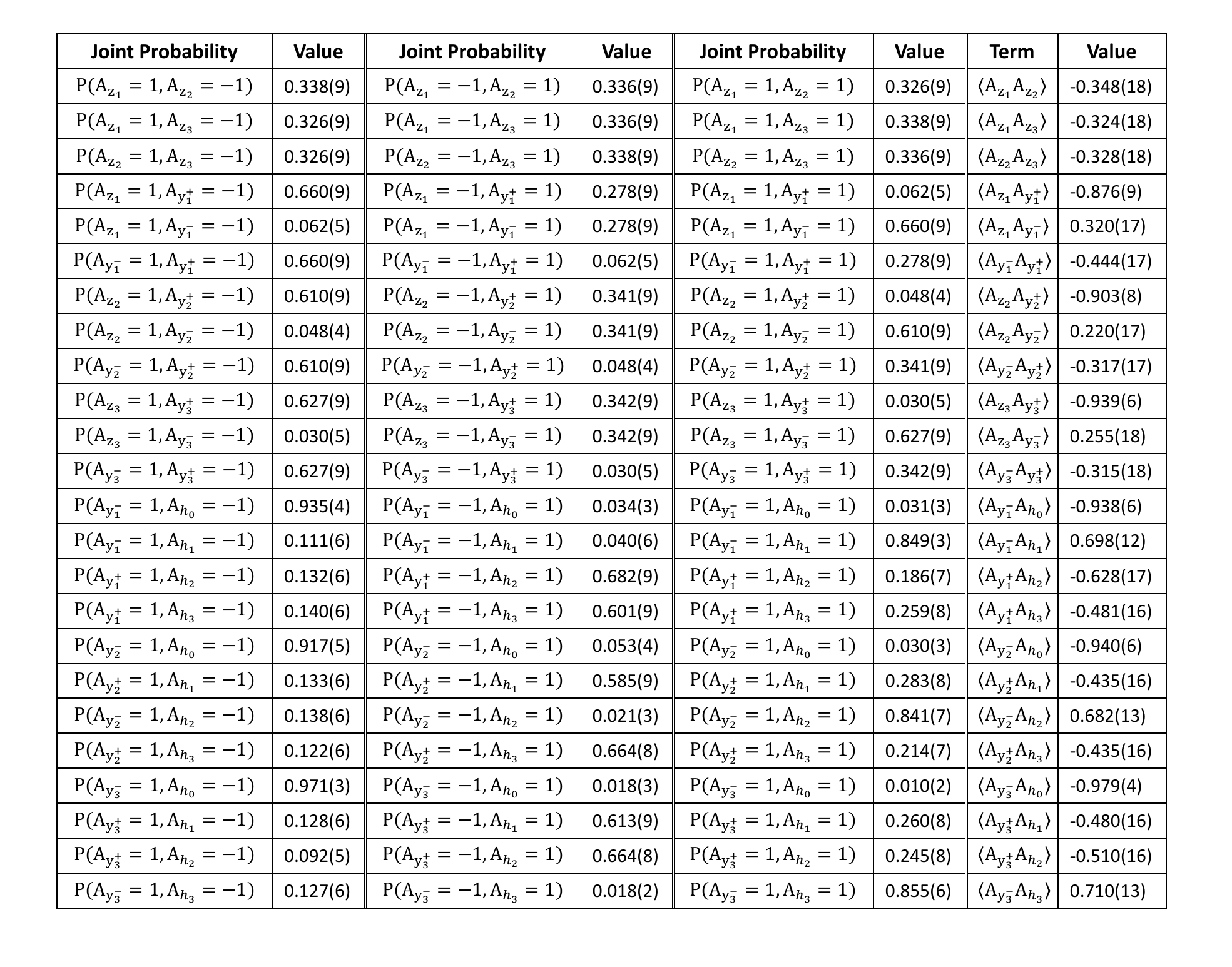}
\caption[Supplementary Table. 3 ]{The measured joint probabilities for all the compatible
pairs $A_{i}$ and $A_{j}$ in the KS inequality under the input state $|\Psi_7\rangle = \frac{1}{\protect\sqrt{3}} (|0\rangle
+|1\rangle +|2\rangle)$}
\end{table}

\begin{table}[tbp]
\includegraphics[width=18cm,height=24cm]{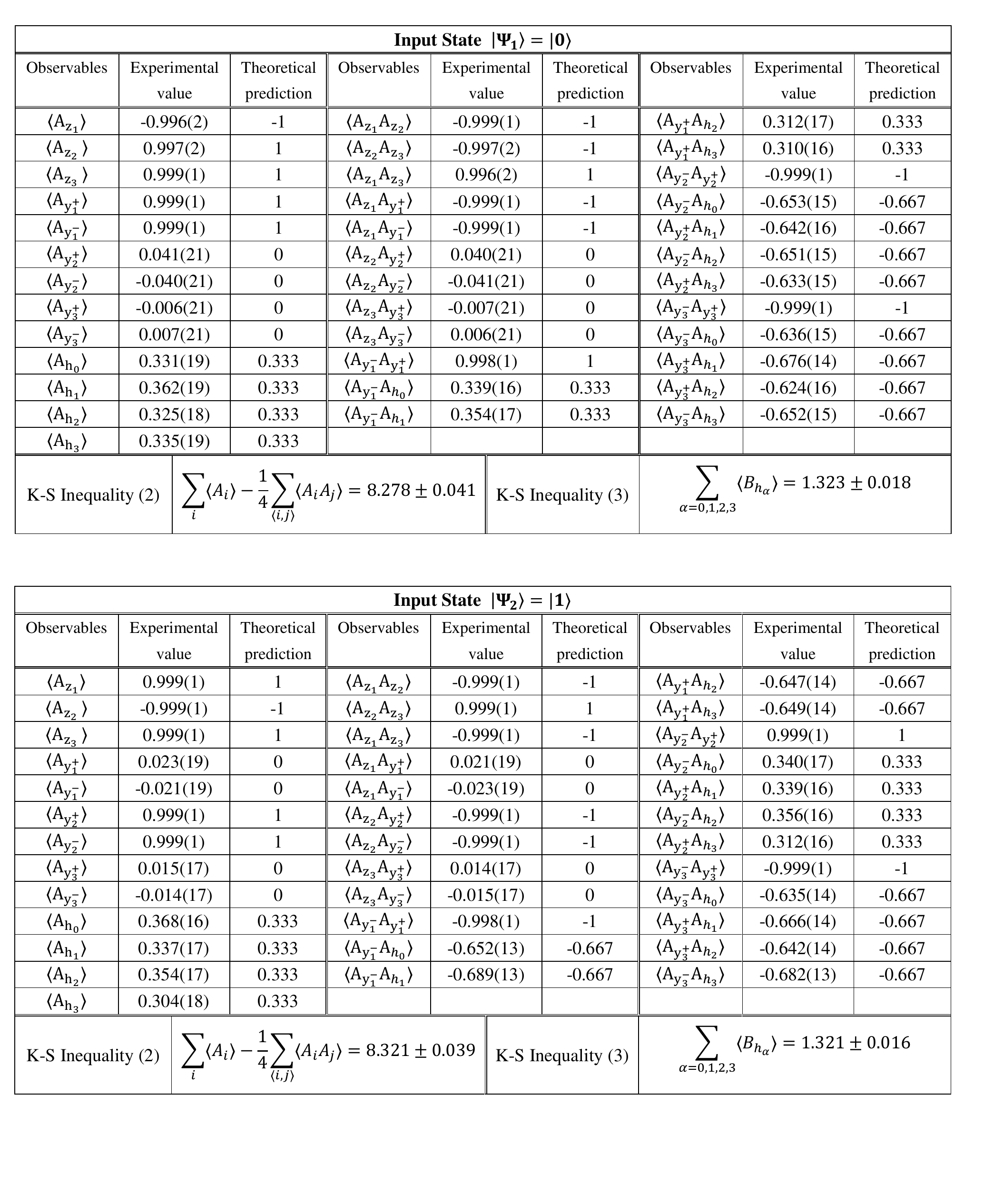}
\caption[Table.4_1]{The measured expectation values $\left\langle A_{i}\right\rangle $ and
the correlations $\left\langle A_{i}A_{j}\right\rangle $ for all the compatible pairs under
input states $|\Psi _{1}\rangle =|0\rangle $ and $|\Psi _{2}\rangle =|1\rangle $.}
\end{table}

\begin{table}[tbp]
\includegraphics[width=18cm,height=24cm]{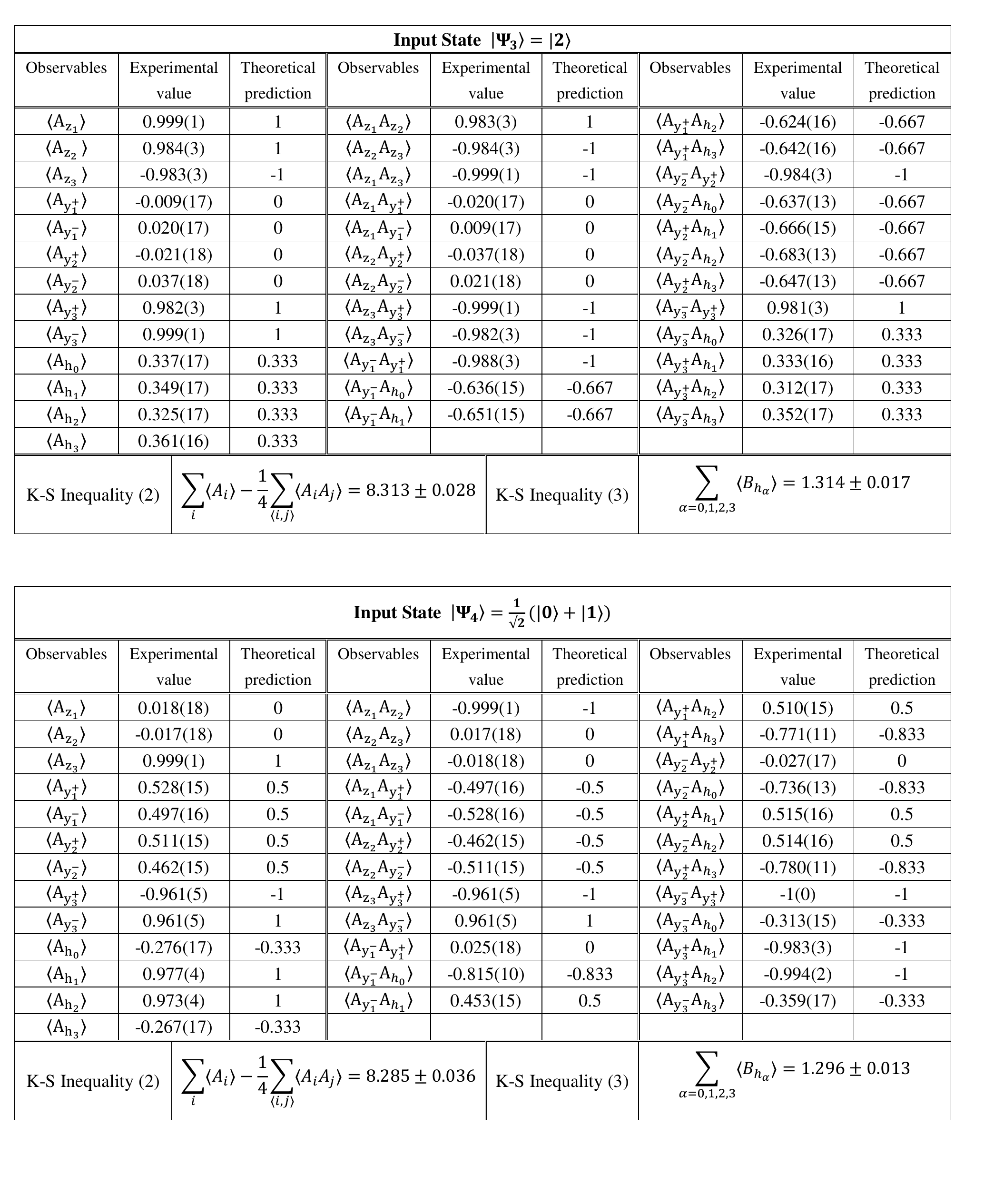}
\caption[Table.4_2]{The measured expectation values $\left\langle A_{i}\right\rangle $ and
the correlations $\left\langle A_{i}A_{j}\right\rangle $ for all the compatible pairs under
input states $|\Psi _{3}\rangle =|2\rangle $ and $|\Psi _{4}\rangle =(|0\rangle
+|1\rangle )/\sqrt{2}$.}
\end{table}

\begin{table}[tbp]
\includegraphics[width=18cm,height=24cm]{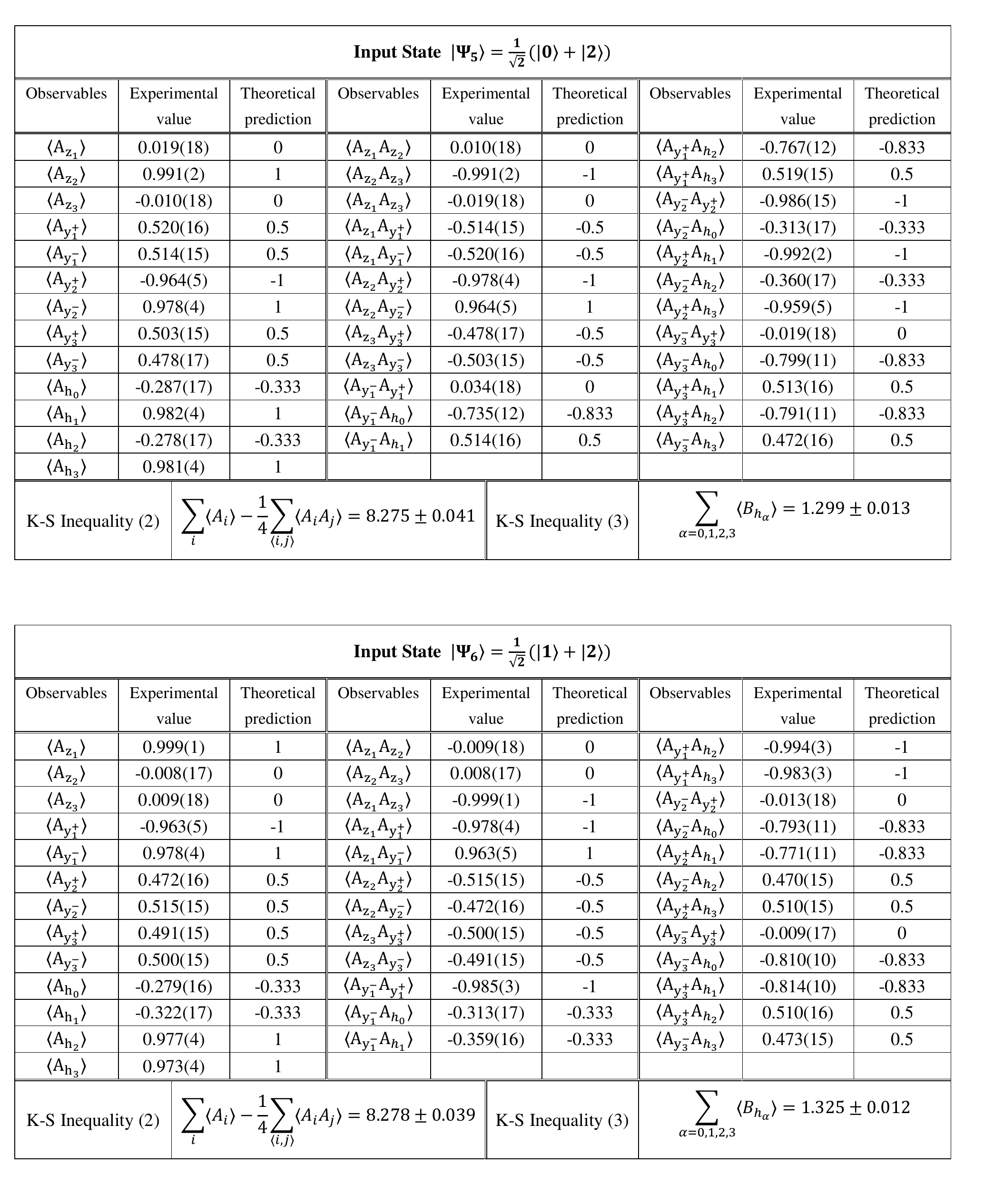}
\caption[Table.4_3]{The measured expectation values $\left\langle A_{i}\right\rangle $ and
the correlations $\left\langle A_{i}A_{j}\right\rangle $ for all the compatible pairs under
input states $|\Psi _{5}\rangle =(|0\rangle +|2\rangle )/\sqrt{2}$ and $|\Psi _{6}\rangle
=(|1\rangle +|2\rangle )/\sqrt{2}$.}
\end{table}

\begin{table}[tbp]
\includegraphics[width=18cm,height=24cm]{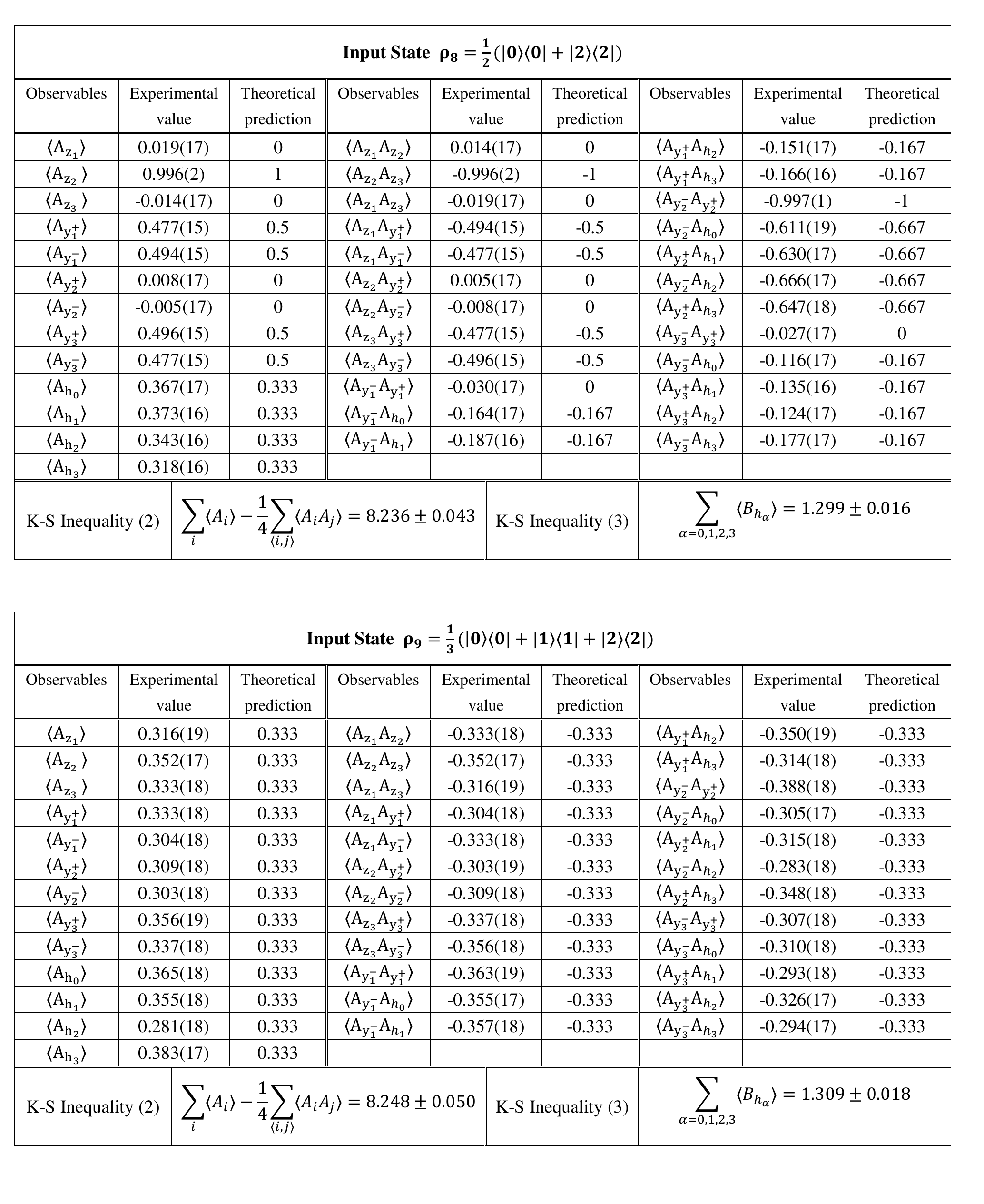}
\caption[Table.4_4]{The measured expectation values $\left\langle A_{i}\right\rangle $ and
the correlations $\left\langle A_{i}A_{j}\right\rangle $ for all the compatible pairs under
input states $\rho _{8}=(|0\rangle \left\langle 0\right\vert +|2\rangle \left\langle
2\right\vert )/2$ and $\rho _{9}=(|0\rangle \left\langle 0\right\vert
+|1\rangle \left\langle 1\right\vert +|2\rangle \left\langle 2\right\vert )/3
$.}
\end{table}

\end{widetext}
\end{document}